\documentclass{appolb}
\usepackage{graphicx}
\usepackage{amssymb}
\usepackage{float}
\usepackage{hyperref}
\usepackage{url}
\usepackage{subfig}

\begin{document}
\title{Evolution of Temperature Fluctuation in a Thermal bath and, its
implications in Hadronic and Heavy-Ion Collisions%
\thanks{Presented at the XI Workshop on Particle Correlations and Femtoscopy by R. Sahoo (Raghunath.Sahoo@cern.ch)}%
}
\author{Trambak Bhattacharya, Prakhar Garg, Raghunath Sahoo 
\address{Discipline of Physics, School of Basic Sciences, Indian Institute of
Technology Indore, Khandwa Road, Simrol, M.P - 452020,
India.}
\\
\vskip 1.5em
{Prasant Samantray}
\address{Centre of Astronomy, School of Basic Science, Indian Institute of Technology Indore, Khandwa Road, Simrol, MP-452020, India}
}
\maketitle
\begin{abstract}
The evolution equation for inhomogeneous and anisotropic temperature fluctuations inside a medium
is derived within the ambit of Boltzmann Transport Equation. Also, taking some existing realistic
inputs we have analyzed the Fourier space variation of temperature fluctuation for the medium
created after heavy-ion collisions. The effect of viscosity on the variation of fluctuations is investigated.
Further, possible implications in hadronic and heavy-ion collisions are explored.
\end{abstract}
  
\section{Introduction}
Fluctuations are commonly discussed for wide range of systems leading to various different phenomena. 
Density fluctuations at all length scales in the second order phase transition,
the event-by-event fluctuation of conserved numbers in heavy ion collisions, important to explore the phase diagram of quantum chromodynamics (QCD), are
some such examples. Much in the same way as number of particles in a certain
region of a system fluctuates, the everyday examples teach us that the
temperature for physical systems can also fluctuate. The dynamics of the
evolving system dictates the temperature fluctuation until the system comes to
equilibrium and therefore it may vary with time and space during evolution.

\begin{figure}[ht!]
\centering
\includegraphics[scale=0.2]{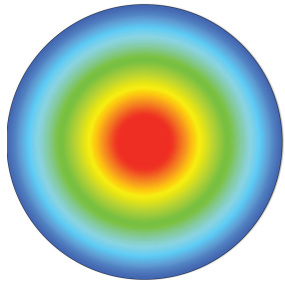}
\caption{{\small Radially varying temperature profile in a medium~\cite{Bhattacharyya:2015nwa}.}}
\label{hotspots}
\end{figure}

Figure~\ref{hotspots} shows a system with radially varying temperature zones.
We model the system to consist of several non-interacting zones whose
temperatures remain constant within a pre-determined time scale. But, after that
given time the temperatures vary. So, the average value as well as the
temperature fluctuation are the quantities which vary with time.

In the present work, the evolution of  fluctuation in inverse temperature
($\beta=1/T$) is discussed for a system resembling with the system formed in
heavy ion collisions using Boltzmann Transport Equation (BTE) in Relaxation Time
Approximation (RTA). It is assumed that the observation time is much less than
the relaxation time of the thermal bath. Further, we extend this study for an
arbitrary observation time.
 
\section{Method}

In order to study the evolution of temperature fluctuation, we consider the
following ansatz for the particle distribution function:
\begin{equation}
 f=e^{-\beta p (1 + \Delta \beta)}.
\end{equation}
Here, we consider a medium with Boltzmann distribution of massless particles
with average inverse temperature $\beta(t)$, at some time slice. To include
anisotropic and inhomogeneous fluctuation we add a function 
$\Delta\beta(\vec{r},\hat{p};t)$, where $\hat{p}$ is an unit vector along the
direction of motion of particle. We use the BTE to calculate the temporal
evolution of $\beta$. Following is the generic form of BTE:
\begin{equation}
\frac{df}{dt}=\frac{\partial f}{\partial t}+\vec{v}.\vec{\nabla} f+\vec{F}.\vec{\nabla}_p f=\mathcal{C}[f].
\label{bte}
\end{equation}
Here $\vec{v}$ is particle velocity, $\vec{F}$ is any external force, $C[f]$ is
the collision term to take care the interaction and $\vec{\nabla}_p$ is the
momentum-space gradient operator. For present study, we assume that the system
experiences no external force, and hence $\vec{F}$ = 0. However, the
inhomogeneity in $\Delta\beta$ still exists. 

\begin{figure}[ht!]
\centering
\subfloat[]{\includegraphics[scale=0.5]{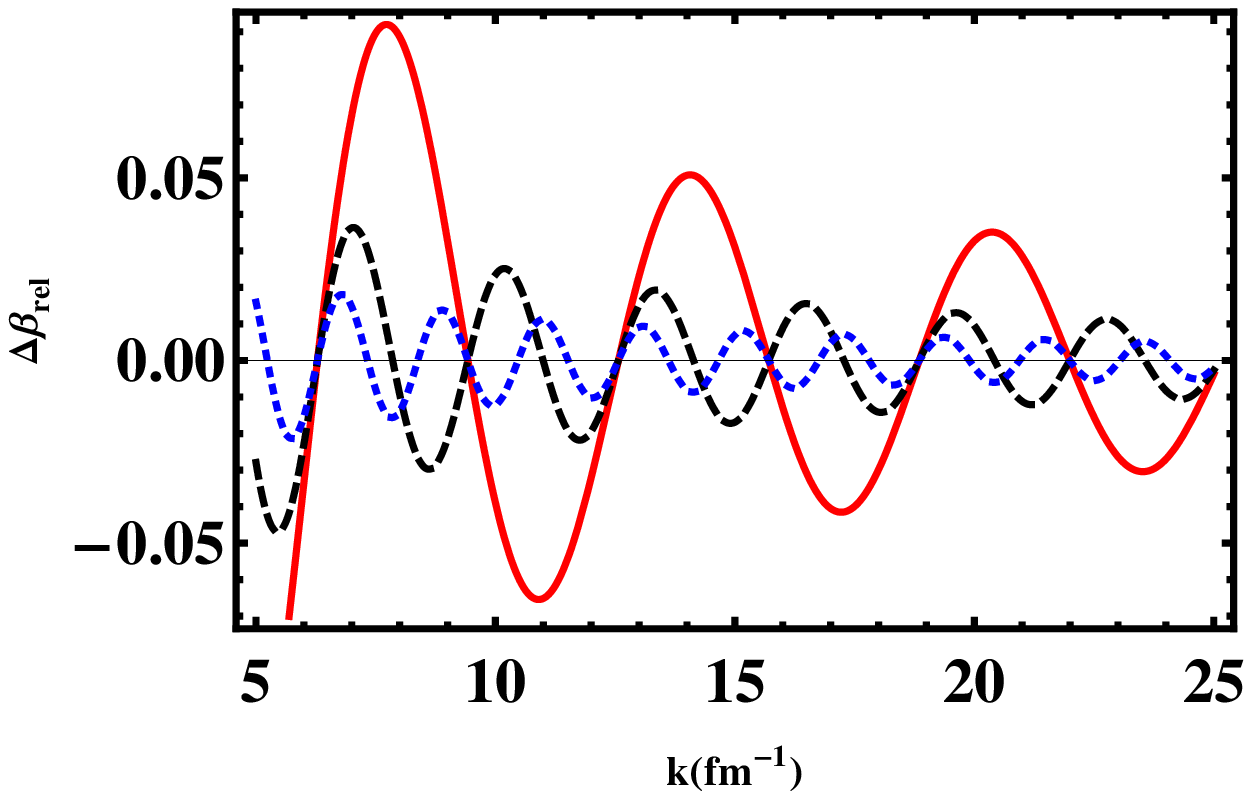}}
\subfloat[]{\includegraphics[scale=0.5]{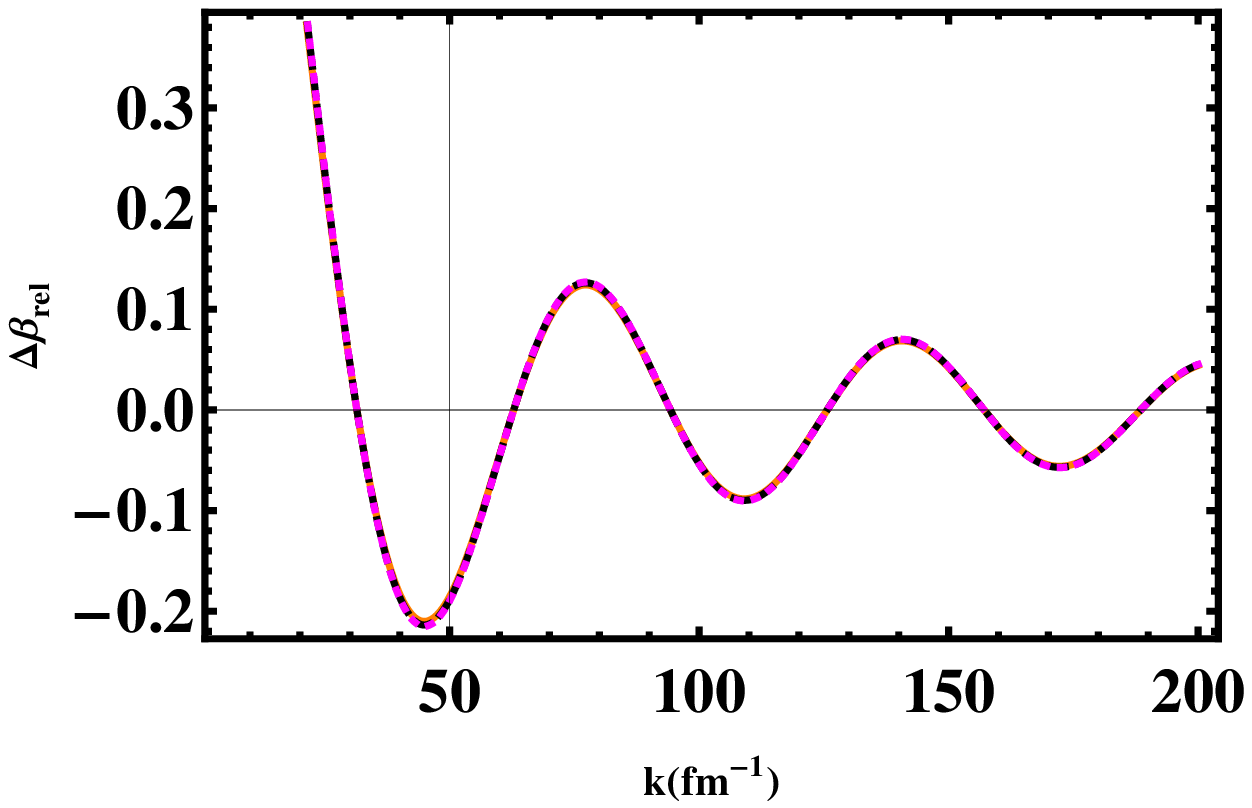}}
\caption{\small Variation of $\Delta \beta_{\mathrm{rel}} (\vec{k};t)$
with $k$. (a)  Red (solid): $(t-t^0) =1 ~fm$, Black (dashed):
$(t-t^0) = 2~fm$ , Blue (dotted): $(t-t^0) = 3~ fm$ for $t_{\mathrm{R}} = 3
~fm$. (b)  Orange (solid): $t_{\mathrm{R}} = 3 ~fm$, Black
(dashed): $t_{\mathrm{R}} = 6 ~ fm$, Magenta (dotted) $t_{\mathrm{R}} = 9~ fm$
for $(t-t^0) = 0.1~fm $~\cite{Bhattacharyya:2015nwa}.}
\label{fig2}
\end{figure}

By assuming an isotropic fluctuation profile and averaging over the whole
solid angle $\Omega$ subtended by $\hat{p}$ the average fluctuations are
derived as follows:

\begin{eqnarray}
\Delta \beta_{\mathrm{av}} (\vec{k};t) &=& \Delta \beta (\vec{k};t^0) e^{-\frac{t-t^0}{t_{\mathrm{R}}}}
\frac{1}{4\pi} \int_{\Omega} e^{-ik\mu(t-t^0)} d\Omega \nonumber\\
&=& \Delta \beta (\vec{k};t^0) e^{-\frac{t-t^0}{t_{\mathrm{R}}}}
\frac{1}{4\pi} \int_{-1}^{1} d\mu e^{-ik\mu(t-t^0)} \int_{0}^{2\pi} d\phi \nonumber\\
\Delta \beta_{\mathrm{rel}} (\vec{k};t) &=& 
\frac{\Delta \beta_{av} (\vec{k};t)}{\Delta \beta (\vec{k};t^0)} \nonumber\\
&=&  e^{-\frac{t-t^0}{t_{\mathrm{R}}}}\frac{\mathrm{sin} k(t-t^0)}{k(t-t^0)}
\label{relfluc}
\end{eqnarray}

here $\vec{k}$ is a constant vector directed along the z-axis. Fig.~\ref{fig2}
shows the parametric Fourier space variation of the $\Delta\beta_{\mathrm{rel}}
(\vec{k};t)$ with time ($t-t^0$) and relaxation time ($t_{\mathrm{R}}$)
respectively. It is observed that the relative fluctuations die down with time.
Additionally, the fluctuations at larger distances towards the periphery of the
medium, are large. Further, we observe no modification of fluctuation with
increasing $t_{R}$ when $(t-t_{0}) << t_{R}$.

To study a more realistic situation, we consider the temperature profiles
\cite{Bhattacharyya:2015nwa} of a evolving medium at different stages of the
evolution of Quark-Gluon Plasma \cite{baiertempprof} created in central
collisions. Given the profile, we get a set of temperatures, their average value
and the temperature fluctuation on top of the average value. We can find out the
variation of the inverse temperature fluctuation at different stages (for
more details see~\cite{Bhattacharyya:2015nwa}).

\begin{figure}[ht!]
\centering
\subfloat[]{\includegraphics[scale=0.5]{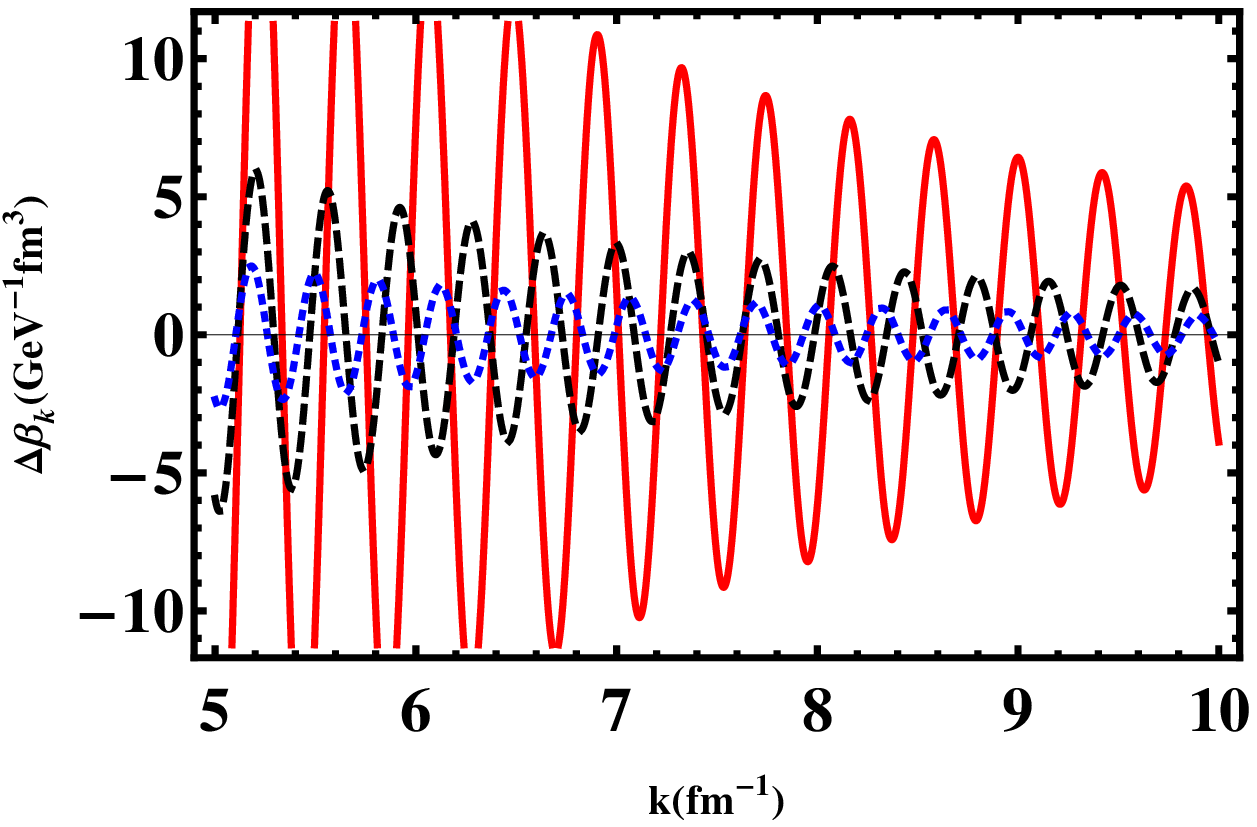}}
\subfloat[]{\includegraphics[scale=0.5]{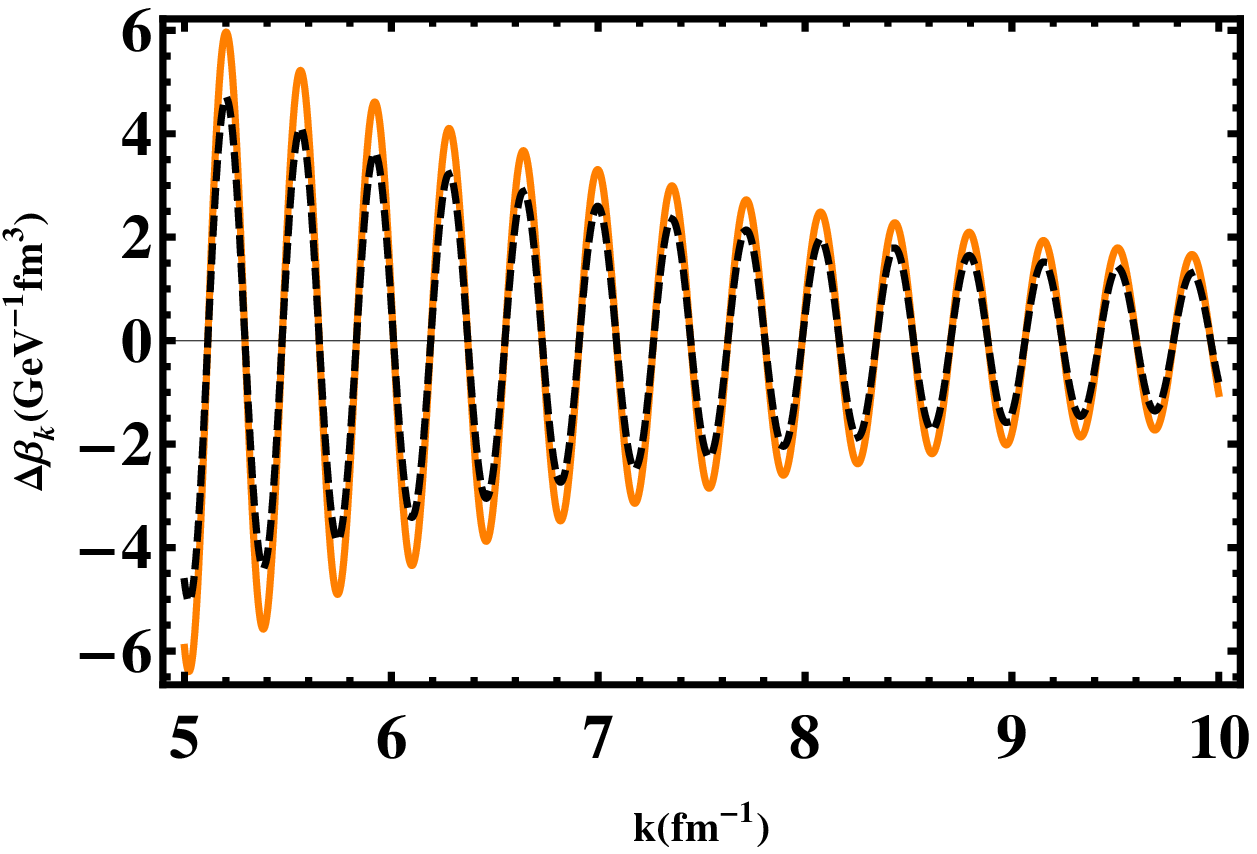}}
\caption{\small Variation of inverse temperature fluctuation in a viscous medium
with $k$. (a) Red(solid):
$\tau=2.2$ fm/c, Black(dashed): $\tau=5.1$ fm/c, Blue(dotted): $\tau=9.1$ fm/c.
$\eta/s=0.08$ for all the figures. (b) Orange(solid): $\eta/s
= 0.08$, Black(dashed): $\eta/s = 0.3$. at $\tau=5.1$ fm/c~\cite{Bhattacharyya:2015nwa}.}
\label{fig3}
\end{figure}

\section{Results and Discussions}
Fig.~\ref{fig3}(a), shows that
the soft modes of $\beta$-fluctuation become dominant at large system radius and
Fig.~\ref{fig3}(b) shows the variation of fluctuation for different viscosities
of the medium. As intuitively expected, higher viscosity favours lower
fluctuations. Within any arbitrary choice of radius shell the relative
fluctuations die down with time. For demonstration, we have chosen the shell
ranging between the radii 14 fm to 15 fm. But our observation remains unaltered
for any other shell.

In view of the connection between the relative temperature fluctuation and the
Tsallis $q$ parameter, as shown in Ref. \cite{wilkprl}, we can compute the
relative temperature fluctuation in QGP produced in a single event existing even
after a long time. The relative temperature fluctuation at the boundary is seen
to be close to the experimentally obtained value 0.018 $\pm$ 0.005 for 0-10\%
central HICs at RHIC with $\sqrt{{s}_{\mathrm{NN}}} =200$ GeV \cite{xuTBW}.


\begin{thebibliography}{99}

\bibitem{Bhattacharyya:2015nwa} 
  T.~Bhattacharyya, P.~Garg, R.~Sahoo and P.~Samantray, arXiv:1510.03154 [hep-ph].

\bibitem{baiertempprof} R. Baier and P. Romatschke, Eur. Phys. Jour. C {\bf 51},
677(2007)

\bibitem{wilkprl} G. Wilk and Z. Wlodarczyk, Phys. Rev. Lett. {\bf 84}, 2770
(2000)

\bibitem{xuTBW} Z. Tang, Y. Xu, L. Ruan, G. v. Buren, F. Wang, and Z.
Xu, Phys. Rev. C {\bf 79}, 051901(2009)

\end{thebibliography}
\end{document}